\documentclass[conference]{IEEEtran}
\IEEEoverridecommandlockouts
\usepackage{cite}
\usepackage{amsmath,amssymb,amsfonts}
\usepackage{tipa}
\usepackage{algorithmic}
\usepackage{graphicx}
\usepackage{textcomp}
\usepackage{xcolor}
\usepackage{url}
\def\BibTeX{{\rm B\kern-.05em{\sc i\kern-.025em b}\kern-.08em
    T\kern-.1667em\lower.7ex\hbox{E}\kern-.125emX}}

\usepackage[most]{tcolorbox}

\tcbset{
  thematicquote/.style={
    colback=gray!10,
    colframe=black,
    boxrule=0.4pt,
rounded corners,  % <- this line changes the border to be rounded
    arc=2mm,
    before skip=10pt,
    after skip=10pt,
    left=6pt,
    right=6pt,
    top=4pt,
    bottom=4pt
  }
}

\begin{document}

\title{Rethinking Code Review Workflows with LLM Assistance: An Empirical Study
} 

\author{
    \IEEEauthorblockN{
        Fannar Steinn A\textipa{D}alsteinsson\IEEEauthorrefmark{1}\IEEEauthorrefmark{2},
       Björn Borgar Magnússon\IEEEauthorrefmark{1}\IEEEauthorrefmark{2},
       Mislav Milicevic\IEEEauthorrefmark{1},\\
       Adam Nirving Davidsson\IEEEauthorrefmark{1},
     Chih-Hong Cheng\IEEEauthorrefmark{2}\IEEEauthorrefmark{3}\thanks{The first two authors contributed equally to this work.}\thanks{Correspondence to: \texttt{chihhong@chalmers.se}}
    }
    \IEEEauthorblockA{\IEEEauthorrefmark{1}WirelessCar Sweden AB, Gothenburg, Sweden}
    \IEEEauthorblockA{\IEEEauthorrefmark{2}Chalmers University of Technology, Gothenburg, Sweden}
    \IEEEauthorblockA{\IEEEauthorrefmark{3}University of Gothenburg, Gothenburg, Sweden}
}

\maketitle

\begin{abstract}

Code reviews are a critical yet time-consuming aspect of modern software development, increasingly challenged by growing system complexity and the demand for faster delivery. This paper presents a study conducted at WirelessCar Sweden AB, combining an exploratory field study of current code review practices with a field experiment involving two variations of an LLM-assisted code review tool. The field study identifies key challenges in traditional code reviews, including frequent context switching, insufficient contextual information, and highlights both opportunities (e.g., automatic summarization of complex pull requests) and concerns (e.g., false positives and trust issues) in using LLMs. In the field experiment,  we developed two prototype variations: one offering LLM-generated reviews upfront and the other enabling on-demand interaction. Both utilize a semantic search pipeline based on retrieval-augmented generation to assemble relevant contextual information for the review, thereby tackling the uncovered challenges. Developers evaluated both variations in real-world settings: AI-led reviews are overall more preferred, while still being conditional on the reviewers' familiarity with the code base, as well as on the severity of the pull request. 

\end{abstract}


\section{Introduction}

Code review (CR) is a cornerstone of modern software engineering, ensuring code quality, defect detection, and team knowledge sharing. Nevertheless, as software systems scale and development cycles accelerate, traditional manual code review practices struggle with inefficiencies, reviewer fatigue, and inconsistent outcomes.
The rapid development of Large Language Models (LLMs)~\cite{achiam2023gpt, yang2024qwen2,team2024gemma} has enabled AI to demonstrate strong capabilities across various software engineering tasks, including code generation and bug detection. 
While their integration into the coding activities has been smooth, the full potential of LLMs in code review remains underexplored.

Through a combination of observational research and field experiments conducted at WirelessCar Sweden AB\footnote{\url{https://www.wirelesscar.com/}} (where some but not all teams are permitted to utilize AI in development), this study investigates how LLMs can be meaningfully integrated into modern code review workflows to improve developer experience and potentially support review efficiency. The research is structured around two core questions: one diagnostic and one exploratory. The first question (RQ1) focuses on understanding current code review practices within the company and identifying opportunities for AI assistance, while the second question (RQ2) explores how developers perceive and interact with two variations of LLM-based tools during review tasks.

To address these research questions, we conducted a two-phase empirical study (Sec.~\ref{sec:methodology}) at WirelessCar Sweden AB. 

\begin{itemize}
\item Phase~1 involved a field study through semi-structured interviews to uncover practical challenges in existing code review workflows and to identify potential areas where AI assistance could be beneficial. Importantly, based on these findings, we designed and implemented two variations of LLM-assisted review tools, one being an AI-led co-reviewer and the other an interactive assistant, integrated with retrieval-augmented generation (RAG) to provide contextual support. 

\item In Phase~2, we evaluated these tools in a real-world field experiment involving practicing developers. The results show that while the AI-led mode was generally preferred, especially for large or unfamiliar pull requests, preferences were context-dependent. Participants valued the tool's ability to provide faster understanding, improved thoroughness, and helpful contextual insights, though issues such as trust, false positives, and interface limitations were noted.

\end{itemize}

Altogether, our findings (Sec.~\ref{sec:results}) suggest that LLMs can meaningfully augment, rather than replace, human reviewers, and highlight the importance of adaptive integration strategies that respect developers’ workflow preferences and domain knowledge. Furthermore, our results point to clear design directions for future tools: AI assistance should be seamlessly embedded into existing developer environments (e.g., GitHub, IDEs, Slack), offer concise and structured feedback with minimal latency, and support both proactive and reactive interaction modes. For maximal utility, the assistant must be context-aware and capable of leveraging relevant project artifacts like code diffs, source files, and requirement tickets. Notably, participants also envisioned the assistant being valuable during the pre-review phase, helping authors improve pull requests before submission. These practical insights inform how LLMs can be responsibly and effectively integrated into real-world software development workflows.

\section{Related Work}

Since the emergence of ChatGPT, large language models have seen widespread adoption in all software engineering tasks, such as code generation, test case creation, or documentation. We refer readers to a review paper~\cite{hou2024large} published in 2024 for a summary of recently conducted research activities. In particular, recent works~\cite{Tufano2024DeepLC,Lin2024,Vijayvergiya2024,Rasheed2024,Alami2025HumanAM,tufano2021towards} have explored the application of AI, in specific LLMs, to automate or support code review tasks. The early result from Tufano~\emph{et al.}~\cite{tufano2021towards} presents a deep learning model trained to replicate reviewer-suggested code changes, aiming to partially automate the code review process. In the era of LLMs, the work from Tufano~\emph{et al.}~\cite{Tufano2024DeepLC} evaluates the measurable impact of AI-generated code reviews in controlled experiments over code with injected issues and code smells. It focuses on the performance of the LLM itself and the quantifiable effects on issue detection, time spent, and reviewer behavior. The work from Alami~\emph{et al.}~\cite{Alami2025HumanAM} performs qualitative, interview-based exploration of developers’ emotional and cognitive responses to AI-provided feedback compared to human-provided feedback in code reviews.  The work from Vijayvergiya~\emph{et al.}~\cite{Vijayvergiya2024} showcases the deployment and evaluation of a large-scale automated system that enforces coding standards and code smells using LLMs in code reviews. The work from Lin~\emph{et al.}~\cite{Lin2024} studies how to fine-tune LLM for better issue detection in code reviews. Similarly, the work from Rasheed~\emph{et al.}~\cite{Rasheed2024} develops a multi-agent LLM-based system for performing autonomous code reviews, focused on technical accuracy, issue detection, and actionable suggestions.

Summarizing the above results, one of the critical gaps is the study of the \emph{preferred interaction} for an engineer assigned with the code-review task in collaboration with an LLM-enabled review assistant, which leads to the primary focus of our work (RQ2). We focus on understanding how LLMs can support, not replace, reviewers via practical integration strategies. We do not consider LLMs as human replacements mainly due to the fundamental concern that LLMs are still prone to hallucinations~\cite{huang2025survey}. Another differentiator is the introduction of semi-structured interviews and thematic analysis to understand the real potential of LLMs in code reviews, reflecting the pain points of an organization developing complex software (RQ1). Understanding the pain points (RQ1) leads to the design of experiments and prototypical LLM-assisted code review tools to assess the preferred interaction (RQ2).

\section{Research Questions}
This study investigates how LLMs can be meaningfully integrated into modern code review workflows to improve developer experience and potentially support review efficiency. The research is structured around two core questions: one diagnostic and one exploratory. The first question focuses on understanding current code review practices and identifying opportunities for AI assistance, while the second question explores how developers perceive and interact with LLM-based tools during review tasks.

\vspace{1mm}

\begin{tcolorbox}[
    colback=gray!10,
    colframe=black,
    arc=10pt,
    boxrule=0.5pt,
    left=5pt,
    right=5pt,
    top=5pt,
    bottom=5pt,
    width=\linewidth
    ]

\textbf{\textit{RQ1}}: \textit{What practices, challenges, and expectations characterize modern code review processes, and where do developers see potential for AI-based assistance?}

\end{tcolorbox}

This question aims to identify how code reviews are currently performed within the company, what challenges developers face, and how AI can be introduced in the code review process, what tasks it can effectively support, and the optimal balance between automation and human involvement.

\vspace{1mm}

\begin{tcolorbox}[
    colback=gray!10,
    colframe=black,
    arc=10pt,
    boxrule=0.5pt,
    left=5pt,
    right=5pt,
    top=5pt,
    bottom=5pt,
    width=\linewidth
    ]

\textbf{\textit{RQ2}}: \textit{How do developers perceive LLM-assisted code review tools, and what is the preferred interaction?}

\end{tcolorbox}

This question explores the qualitative aspects of AI usage, including developer trust, satisfaction, and potential barriers to adoption, by considering different interaction modes. By answering this question, insights are provided into the usability, limitations, and acceptance of AI-assisted code reviews, helping to inform design decisions and future development of such tools.

\section{Methodology}\label{sec:methodology}

\begin{figure}[t]
\centering
\includegraphics[width=\columnwidth]{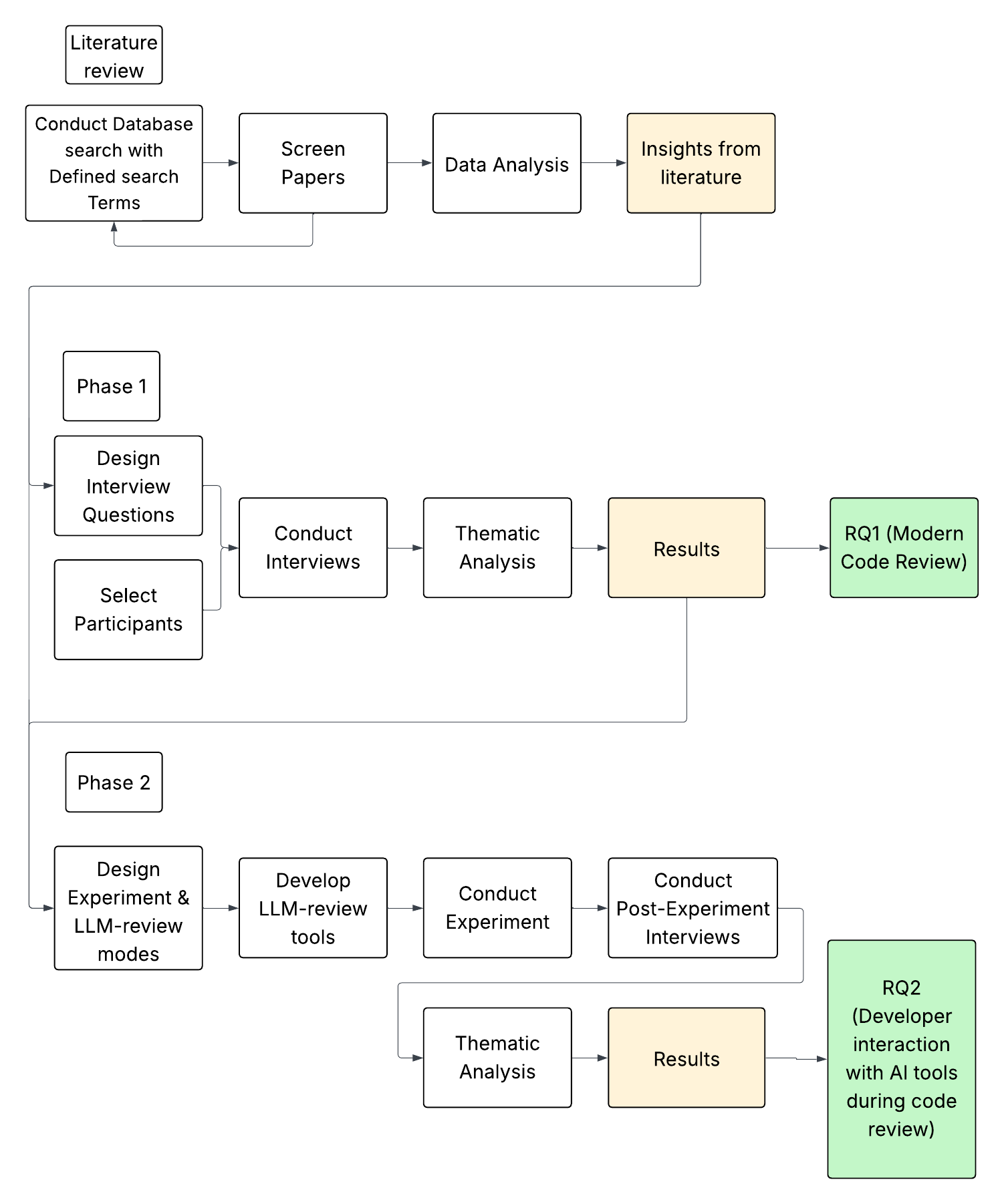}
\vspace{-5mm}
\caption{Flowchart detailing the research workflow}
\vspace{-5mm}
\label{fig:Research_Flowchart}
\end{figure}

Towards the above research questions, grounded in the literature review, we adopt a two-phase research design that is aimed at understanding both the current code review processes in a natural environment and then intervening with an AI-based solution to observe the effects (see Fig.~\ref{fig:Research_Flowchart} for the research flowchart). Phase~1 involves an exploratory case study at WirelessCar to thoroughly understand existing manual code review practices, identify specific pain points, and potential AI application areas. The goal is to identify typical workflow steps, uncover key pain points, clarify why certain inefficiencies arise, how success is evaluated, capture the technical and organizational environment, and where AI might offer possible improvements. 
Phase~2 aims to assess how software developers at WirelessCar experience and evaluate the integration of a software artifact featuring LLM-assisted code review tools into their review process, thereby identifying design implications and best practices.

Importantly, rather than focusing on performance metrics or direct comparisons to existing tools as conducted in prior works, our emphasis on Phase~2 is on capturing qualitative insights into how developers interact with the AI assistant, what kinds of support they find valuable, and how such a tool might be ideally integrated into real-world code review practices. To make such an assessment possible, we created two variations of LLM-enabled review, namely the \emph{AI-led mode (co-reviewer)} and the \emph{interactive mode}. The design of the two modes is based on challenges and improvement opportunities identified during Phase~1 as well as insights collected from the literature.

Due to our research being qualitative in nature, in both phases, semi-structured interviews were conducted to receive feedback, and thematic analysis~\cite{Braun01012006} has been applied to analyze the interview scripts being collected.

\subsection{Phase 1: Understanding the existing manual code review practices (field study)}

In our field study, semi-structured interviews were conducted. Such a format balances consistency where all participants receive a core set of similar questions with flexibility where researchers can ask follow-up or clarifying questions as fits the flow of the interview~\cite{https://doi.org/10.1002/jac5.1441}. The interview questions were developed around domains such as code review processes, challenges in code reviews, measuring code review success, and current and potential AI use cases to cover the previously mentioned objectives. The interviews were designed to fit comfortably within a 30-minute timeframe. However, in practice, the interviews ranged from 15 to 40 minutes each, with most lasting approximately half an hour.

We used \emph{convenience sampling}~\cite{etikan2016comparison} by interviewing people within the WirelessCar who were able and willing to participate. Participants were recruited via announcements on Slack channels and informal Slack messages that announced the study's purpose and form of the interview. Interviewees who handled different parts of the code and were reviewers at varying seniorities were sought. Additionally, some interviewees were from the same development team to gain multiple perspectives from the same team. Despite being limited in randomness, we believe our sampling still enables insights into a relatively wide range of review practices.

\begin{table}[t]
\centering
\caption{Overview of Interview Participants in Phase 1}
\label{table:interviewees.phase.1}
\renewcommand{\arraystretch}{1.3} % Improve row spacing for readability
\begin{tabular}{lll} \hline\hline
\textbf{Participant ID} & \textbf{Role} & \textbf{Team} \\ \hline
P1 & Quality Assurance Specialist & Team A \\ 
P2 & Application Developer & Team A \\ 
P3 & Software Engineer & Team A \\ 
P4 & Software Engineer & Team A \\ 
P5 & Security Engineer& Team B \\ 
P6 & Software Engineer & Team C \\ 
P7 & Software Engineer& Teams D \& B \\ 
\end{tabular}

\vspace{-5mm}
\end{table}

A total of seven participants were interviewed and are listed in Table~\ref{table:interviewees.phase.1}. The interviewees varied in gender and age, and the sample included software engineers, developers, security engineers, and quality assurance specialists across four development teams. Teams ranged in size from under five members to around sixteen members. The interviewees had different areas of expertise, with varying levels of experience and skill. The goal was to include individuals who could discuss both strategic and day-to-day review practices. Data saturation was reached after the seventh interview, as no new information or themes were emerging. This decision is grounded in the principle that qualitative data collection can be considered sufficient when additional interviews fail to yield new insights~\cite{doi:10.1177/1525822X05279903}. Data saturation is defined as the point at which no new themes are observed in the data, and studies show that although saturation often occurs within twelve interviews, the basic elements of major themes are frequently present as early as six~\cite{doi:10.1177/1525822X05279903}. The current sample is somewhat heterogeneous in terms of role and team, and the consistency of responses across participants suggested that the core aspects had been adequately captured. Although an eighth interview had been scheduled, the participant canceled, but given that saturation had already been achieved, it was decided not to reschedule the interview.

All interviews were held in English and conducted in the participant's active work environment, either on-site at WirelessCar's office or remotely via Microsoft Teams (two participants were interviewed remotely). This alignment with genuine working conditions is consistent with a field study approach since it allows developers to reference real pull requests, team communication channels, and examples of ongoing tasks. They could describe, for example, how a large refactoring pull request (PR) or an urgent bug fix impacted their review process in real-time. Conducting interviews in this natural setting helps reinforce the authenticity of participant responses and results in findings consistent with everyday workflows at WirelessCar.

\subsection{Phase 2: Evaluating the LLM-enabled code review (field experiment)}

\subsubsection{Data Collection}
To explore developer experience with the AI assistant during code reviews, two data collection methods were employed to capture user perspectives and behavioral interaction patterns. The primary data source consisted of post-interaction interviews with each participant. These were supported by a secondary data source consisting of researcher observation notes recorded during review sessions. 

Participants were recruited again through convenience sampling, where individuals who participated in the earlier phase of the study were invited to return for Phase~2. However, only five were available and agreed to participate again. To reach a total of ten participants, five additional individuals were recruited via internal Slack channels, where the study's purpose and structure were briefly described. Of the ten participants, four belonged to the team responsible for the pull requests used in the experiment, while the remaining six were from other teams, allowing comparison across varying levels of codebase familiarity. Table~\ref{tab:participants.phase2} provides an overview of the participants, including their roles and team affiliations. The code used in the experiment originated from Team A, which is the team associated with the familiar participants. 

\begin{table}[t]
\centering
\caption{Overview of Experiment and Interview Participants in Phase 2}
\label{tab:participants.phase2}
\renewcommand{\arraystretch}{1.3}
\begin{tabular}{lll} \hline\hline
\textbf{Participant ID} & \textbf{Role} & \textbf{Team} \\ \hline
P1 & Quality Assurance Specialist & Team A \\ 
P2 & Application Developer & Team A \\ 
P3 & Software Engineer & Team A \\ 
P5 & Security Engineer & Team B \\ 
P7 & Software Engineer& Teams D \& B \\ 
P8 & Software Engineer & Team E \\ 
P9 & Software Engineer & Team A \\ 
P10 & Software Engineer & Team E \\ 
P11 & Software Engineer & Team F \\ 
P12 & Software Engineer & Team G \\ 
\end{tabular}
\vspace{-5mm}
\end{table}

\subsubsection{Experiment Setup}\label{subsec:experiment_setup}

The field experiment is designed to evaluate the developers' experiences with LLM-based code review assistance under two distinct interaction modes that were selected based on the findings from the Phase 1 field study. As indicated by the interview data in that earlier phase, developers most frequently identified a need for clearer up-front summaries of code changes as well as on-demand explanations for specific architectural or contextual details.

\begin{figure}[t]
\centering
\includegraphics[width=\columnwidth]{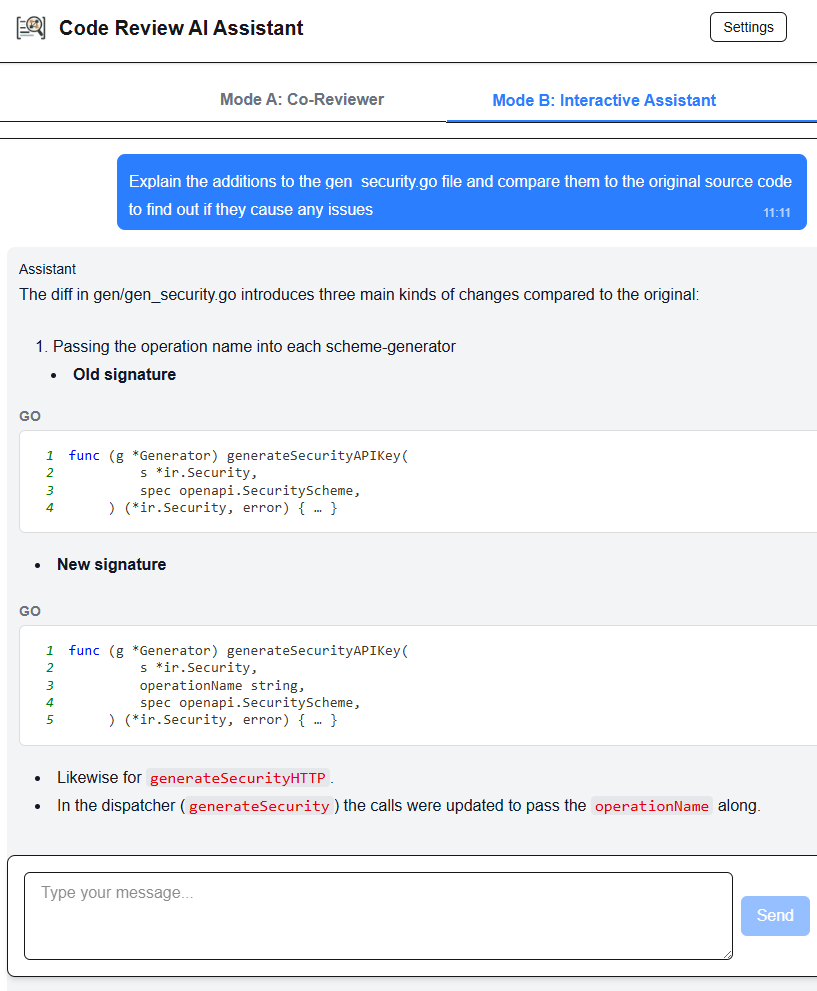}
\caption{Screenshot of the LLM-assisted code review interface in Mode B, reviewing a pull request from an open-source project available at \texttt{https://github.com/ogen-go/ogen/pull/1440}.}
\vspace{-3mm}
\label{fig:ui_screenshot_1}
\end{figure}

In each iteration of the experiment, a single participant engaged in a traditional code review scenario, conducted within their familiar development environment using their usual tools and platforms. In addition to these standard resources, the participant was provided access to our created LLM code review assistant as illustrated in Fig.~\ref{fig:ui_screenshot_1}. The task assigned to the participant was to perform two code reviews, each on a different pull request from two separate repositories within the company. The participant followed a different interaction mode with the AI assistant for each review. All sessions took place in conditions that mirrored each participant's normal working situation as closely as possible.

The two interaction modes are described below:

\begin{itemize}
    \item \textbf{Mode A: Co-Reviewer} – In this mode, the AI assistant automatically generated a summary of the code under review, highlighting major changes and any potential points of interest or concern, before the reviewer started their own examination. The reviewer could then use this information to guide their review, optionally asking the AI-assistant follow-up questions. The participant could query the AI for clarification or more details about the summarized areas. This mode directly targeted the challenge previously identified as lacking immediate context, particularly for large or complex PRs.
    
    \item \textbf{Mode B: Interactive Assistant} – In this mode, the reviewer reviewed the code in their typical manner. The AI assistant did not proactively provide a summary or suggest issues upfront. Instead, the reviewer was free to consult the AI on demand, requesting clarifications about specific parts of the code or asking higher-level architectural questions. Reviewers were encouraged to ask targeted queries rather than requesting an overall summary of the changes. Here, the AI assistant only responded when explicitly prompted. This design choice not only reflected the Phase 1 feedback on needing a lightweight, on-demand tool but also addressed an identified challenge in related studies~\cite{Tufano2024DeepLC}, where automatically highlighted lines can cause reviewers to miss other important areas. By making the AI passive, reviewers maintained their usual workflow and examined the entire codebase without unconsciously depending on the AI’s initial hints. 
    
\end{itemize}

To minimize review variability while still enabling comparison across interaction types, two specific pull requests of similar size and complexity were selected from the WirelessCar codebase. The selected pull requests each involved a moderate amount of change, requiring genuine reviewer effort without being excessively large or trivial. Before the first experiment session, a pilot run was conducted with two internal developers who were not part of the main participant pool. The pilot involved a walkthrough of the tool, the review task, and the interaction modes, followed by an informal test of the tool using the selected pull requests. The pilot was used to assess the suitability of the selected PRs, detect any major usability or technical issues in the tool, and evaluate whether the introduction and guidance provided to participants were sufficiently clear and effective. Each participant in the actual experiment then reviewed both PRs, using Mode A for one PR and Mode B for the other. The assignment of modes to PRs was rotated across participants to mitigate ordering effects. By having all participants conduct the same two code reviews with alternating modes, this approach allowed for a more controlled comparison of interaction styles while ensuring the tasks remained realistic and relevant to actual code review practices.

To investigate how varying levels of codebase familiarity may influence the use of the AI assistant, both developers belonging to the team that owns the selected PRs and developers from unrelated teams were invited to participate. Phase 1 interviews indicated that limited contextual knowledge can have negative effects on the review quality, and therefore, measuring differences in AI reliance across these two participant groups was expected to produce further insights.

Before starting the code review sessions, each participant was given a short onboarding briefing to explain the tool's features as well as the structure and setup of both the study and the experiment. The two different interaction modes were introduced, along with guidance on how to effectively prompt the AI assistant to obtain useful and relevant responses. The participant then completed the two review sessions consecutively. While no in-depth feedback or direct assistance was provided during the sessions, limited guidance was offered when needed, for example, if participants inquired about specific ways of interacting with the assistant. Participants were also encouraged to think aloud during the experiment, often verbalizing their reasoning, confirming the AI’s suggestions, or commenting on its usefulness. If the assistant’s response did not meet expectations, participants were occasionally guided to try rephrasing or retrying the query.  Researchers were present to observe the sessions, record notes, and perform the post-session data collection. Following the review sessions, short semi-structured interviews were conducted with the participant to reflect on their experience across the two modes and in comparison to their regular code review workflow.

\subsubsection{Artifact Implementation}\label{subsec:artifact_implementation}

The software artifact developed for this study was a web-based chat interface designed to explore and evaluate different interaction styles in AI-assisted code reviews. Its primary purpose was to enable structured experimentation by allowing researchers to observe how developers interact with AI assistance in different contexts. The source code for the artifact is freely available (under GPLv3) on GitHub: frontend\footnote{\url{https://github.com/BearPays/code-review-assistant-ui}} and backend\footnote{\url{https://github.com/BearPays/code-review-assistant-back}}. While not intended to be a production-level system, the artifact was designed to be realistic and usable enough to engage developers meaningfully. It supported live interaction through a chat interface, maintained session-specific context, and allowed researchers to configure session parameters such as interaction mode and PR data source.

\begin{figure}[t]
\centering
\includegraphics[width=0.8\columnwidth]{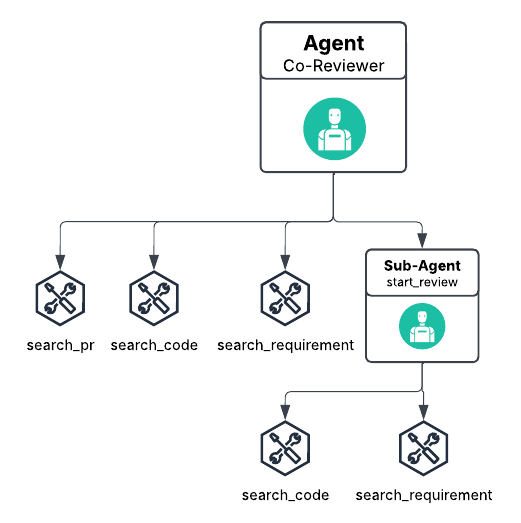}
\caption{Agentic tool structure in Co-Reviewer mode.}
\vspace{-5mm}
\label{fig:co_reviewer_agent}
\end{figure}

The artifact consisted of a chat interface connected to a backend system that integrated OpenAI's \texttt{o4-mini}\footnote{\url{https://platform.openai.com/docs/models/o4-mini}} language model via API. The artifact was also supported by a Retrieval Augmented Generation (RAG) infrastructure built with LlamaIndex\footnote{\url{https://www.llamaindex.ai/}}. This setup enabled the AI assistant to produce more informed and context-aware responses by using project data such as code diffs, related source code files, and associated feature requirements (Jira tickets). The RAG index had to be manually prepared and indexed before experiments, ensuring complete control over the data available to the model in each experimental session. As highlighted by the literature and the Phase~1 interview results (see Sec.~\ref{subsec:code_review_challenges}), lacking a broader repository context can lead to superficial AI feedback that overlooks critical design or architectural concerns. This RAG setup ensured that the LLM can reference deeper project-level information on demand. This setup not only enhanced the AI assistant’s capacity to generate context-aware suggestions but also directly targeted the gap in existing tools and recent studies on the subject.

Internally, the assistant interacts with three core semantic tools:

\begin{itemize}
    \item \texttt{search\_pr}: accesses PR diffs and metadata.
    \item \texttt{search\_code}: provides the full, unmodified source code of the repository.
    \item \texttt{search\_requirements}: contains the feature requirement (the Jira ticket) motivating the PR.
\end{itemize}

In Mode A (Co-Reviewer), a fourth tool, \texttt{start\_review}, was added. This tool contained a sub-agent that was designed to perform an initial, structured code review based on the full PR data and guided by a detailed review-specific prompt. Unlike the main agent, this sub-agent did not use \texttt{search\_pr}, as all PR data was injected into its initial context via a prompt. This ensured that the agent considers everything in the PR data and examines each file change. By retrieving the PR data via a query engine, the agent might not consider all the data as required when generating a complete code review of all changes. The agentic structure for this setup is shown in Fig.~\ref{fig:co_reviewer_agent}.

\section{Results}\label{sec:results}

\subsection{Results from Phase 1}\label{subsec:results_from_phase_1}

Six themes emerged from the thematic analysis of the qualitative data and can be seen in Table~\ref{tab:themes.phase.1}. 

\begin{table*}[h]
\centering
\caption{Identified themes and their descriptions from analysis of interview data in Phase~1.}
\label{tab:themes.phase.1}
\renewcommand{\arraystretch}{1.3} % Improve row spacing for readability
\begin{tabular}{lp{11cm}} \hline\hline
\textbf{Theme} & \textbf{Description} \\ \hline
Informal Review Process and Practices & 
Describes how teams coordinate and manage code reviews in practice, including informal communication, tool use, and the absence of structured processes or metrics. \\
Review Strategies and Evaluation Focus & Captures what developers focus on during the actual code review process. \\
Learning, Knowledge, and Review Expertise & Explores how code reviews serve as opportunities for learning and knowledge sharing within teams. It also captures the role of reviewer expertise in conducting effective reviews and the challenges that arise when reviewers lack sufficient understanding of the codebase or architecture. \\
Code Review Challenges & Identifies recurring challenges and inefficiencies encountered in the code review process. \\
Current AI adoption & Describes the current state of AI tool usage in development and code review. \\
Possible AI adoption in code reviews & Describes developers’ expectations, suggestions, and concerns regarding future AI assistance in code reviews. \\
\end{tabular}
\vspace{-4mm}
\end{table*}

\begin{table*}[h]
\centering
\caption{Identified themes and their descriptions from analysis of data from Phase 2.}
\label{tab:phase2_themes}
\renewcommand{\arraystretch}{1.3} % Improve row spacing for readability
\begin{tabular}{lp{11cm}} \hline\hline
\textbf{Theme} & \textbf{Description} \\ \hline
Accuracy, Reliability, and Trust & Focuses on the perceived correctness of AI-generated feedback, concerns about over-reliance, and varying levels of trust in the assistant’s recommendations. \\
Efficiency and Thoroughness & Captures how the assistant affects review speed, cognitive load, issue detection, and the overall thoroughness of the code review process. \\
Integration Expectations and Limitations & Highlights developer expectations for seamless integration, responsive design, and context-aware suggestions, while also surfacing frustrations related to current UX and tooling limitations. \\
\parbox[t]{5.3cm}{Usage Contexts and \\ Interaction Patterns} & 
Describes how interaction with the assistant varied based on review context, including preferences for different modes, alternative usage strategies, and team-specific practices. \\
\end{tabular}
\vspace{-2mm}
\end{table*}

\subsubsection{Observed Code Review Process}\label{subsec:wirelesscar_code_review_process}

When reviewing the \textit{informal review process at WirelessCar and their practices}, it appears to be similar to the typical asynchronous, tool-supported nature of modern code review (MCR) practices. However, the reliance on informal assignment and communication sometimes through Slack introduces variability, which might contrast with the structured practice of assigning a specific reviewer to the code patch commonly found in MCR practices. Additionally, WirelessCar places significant emphasis on contextual understanding and relies heavily on team members with domain expertise to assess the critical components. This contrasts with the broader responsibility-sharing model commonly found in large organizations that apply MCR.\footnote{Due to space limits, quotes collected from the interview as well as detailed findings are not listed.
}

\subsubsection{Common Challenges in Code Reviews}\label{subsec:code_review_challenges}

One of the most frequently mentioned challenges across the developer interviews was the issue of \textbf{delayed reviews}. Situations were described where PRs remained unreviewed for extended periods, often requiring repeated reminders for a reviewer to take action. One interviewee described it as:

\begin{tcolorbox}[thematicquote]
\itshape``Sometimes you need to ping people more often, and sometimes the PR is very big, so people don't dare to pick it up'' \hfill [P2]
\end{tcolorbox}

Several interviewees reported difficulties regarding reviewing \textbf{large or complex PRs}. PRs that combine new features, refactoring, and changes to infrastructure often become overwhelming. They note that this can result in more superficial reviews or longer delays.

\textbf{Context switching} was also noted as another major challenge. Developers mentioned that the cognitive burden of pausing ongoing work to review code, not at all related to their current work, to be challenging. This requires additional time to regain focus, and one interviewee highlighted the time lost cause of this:

\begin{tcolorbox}[thematicquote]
    \itshape``As soon as you need to context switch, even if it’s just a three-minute thing, it’s 20 minutes of lost time.'' \hfill [P7]
\end{tcolorbox}

Several interviewees explained that they sometimes \textbf{lack sufficient context when reviewing the code}. Sometimes, important details about why the change was made or its expected impact are missing from the PR description. This increases the time it takes for the reviewer to comprehend the PR and effectively point out defects or problems.

\subsubsection{Current Use of AI in Software Development}\label{subsec:current_use_of_AI}

The interviews revealed that common AI tools like GitHub Copilot\footnote{\url{https://github.com/features/copilot}} and ChatGPT\footnote{\url{https://chatgpt.com/}} are commonly used for development tasks. Interviewees mentioned tasks like writing boilerplate code, assisting with syntax, and quickly generating documentation. One example mentioned was:

\begin{tcolorbox}[thematicquote]
\itshape``[...] as to help to create the boilerplate stuff, it's outstanding, right? I mean, you do it in 30 seconds instead of a couple of hours. So I try to use it as much as possible during the development process.'' \hfill [P7]
\end{tcolorbox}

However, not all teams are allowed to use AI-generated code or share information (such as source code) with AI tools. Furthermore, the teams that are allowed to use AI tools only utilize them via an enterprise subscription, where the data provided is not used for training the models utilized by the tools. This ensures that no data is leaked from the organization via the use of AI tools.

All interviewees who use AI tools reported positive experiences in their development work. 
\textbf{None of the interviewees reported using AI tools as part of the formal code review process}. One interviewee did mention that some reviewers might occasionally paste code into tools like ChatGPT for clarification or explanation during reviews. However, beyond such informal use, AI has not been formally integrated into the code review workflow in any of the interviewed teams.

\subsubsection{Potential AI Use Cases in Code Review}\label{subsec:potential_use_for_AI_in_code_reviews}

The interviewees expressed interest in potential AI integrations with the code review processes. They mentioned features like summarizing PR changes and accompanying descriptions where AI could generate a concise summary and help reviewers to quickly understand the intent of a code change. One interviewee mentioned:

\begin{tcolorbox}[thematicquote]
\itshape``When you create the pull request, an AI bot could say, 'Hey, you're trying to achieve this—do you want this as your summary or description?'' \hfill [P5]
\end{tcolorbox}

Interviewees also mention possibilities such as AI assisting in validating whether code meets stated requirements. Additionally, interviewees highlighted the potential for AI to detect hidden bugs or vulnerabilities, such as race conditions, dependency issues, or other subtle defects that human reviewers might overlook. As one developer put it:

\begin{tcolorbox}[thematicquote]
\itshape``An AI would probably be able to identify a race condition, for instance, which, as I said, is an example that's borderline impossible to catch on the fly. In three minutes, you're never going to find that.'' \hfill [P7]
\end{tcolorbox}

Finally, some interviewees mentioned potential drawbacks of integrating AI into the code review process. These were mostly concerns about security risks and false positives, which could reduce the reviewer's trust in the AI assistant and divert their attention from real issues. As one interviewee put it:

\begin{tcolorbox}[thematicquote]
\itshape``The problem with those kinds of checks is that, if they're not good enough, you stop reading them. We see that all the time, you get flooded with false positives, and then you miss the real issues because you start ignoring the feedback.'' \hfill [P7]
\end{tcolorbox}

\subsection{Results from Phase 2}\label{sec:results_from_phase_2}

Table~\ref{tab:phase2_themes} provides an overview of the themes along with descriptions of the themes.

\subsubsection{Accuracy, Reliability, and Trust}

Participants frequently commented on the accuracy of the assistant's feedback and how this influenced their trust in the tool. Several interviewees described the assistant as \textbf{generally accurate and capable of identifying relevant issues}. For example, one participant stated:

\begin{tcolorbox}[thematicquote]
\itshape``As I see it, they were quite accurate [...] It was quite nice, not all of them, but a lot of them.'' \hfill [P9]
\end{tcolorbox}

In multiple cases, the assistant’s output was described as \textbf{confirming the developer’s own thoughts or surfacing something they might not have otherwise caught}. Observation notes also reflect this, with one session noting that “the user said the AI caught exactly what he was looking for in a certain file.” In another session, the reviewer remarked that “the summary gave something that he would not have seen.”

Several participants also reported instances where the assistant produced \textbf{incorrect or unclear suggestions}. One participant questioned whether the tool was even “doing what it was asked,” while another described the assistant incorrectly flagging a missing import. One participant noted:

\begin{tcolorbox}[thematicquote]
\itshape``Sometimes it says some slightly strange things.'' \hfill [P8]
\end{tcolorbox}

When asked about concerns, participants expressed different perspectives. Several participants warned of the risk of over-relying on the assistant, especially in Mode A, where the assistant led the review:

\begin{tcolorbox}[thematicquote]
\itshape``It feels like I might get a bit colored by getting the improvements from the LLM [...] I feel like maybe I could miss something else, because I would focus on those improvements a lot.'' \hfill [P3]
\end{tcolorbox}

While some view using AI tools as risky, not all participants saw the assistant as risky. Some viewed it as a low-stakes addition to the workflow, especially when \textbf{its use remained optional}, even if it is not always accurate:

\begin{tcolorbox}[thematicquote]
\itshape``As long as there's an opt-out option, there's no real harm in it.'' \hfill [P5]
\end{tcolorbox}

\begin{tcolorbox}[thematicquote]
\itshape``If we miss 10 [issues] today, we might miss two with a tool like this.'' \hfill [P9]
\end{tcolorbox}

The \textbf{role of trust} emerged as a key factor in shaping how participants viewed the tool. A few emphasized that for the tool to be useful or even used in general, it must be trusted, but not blindly. Misplaced trust in the tool could lead to wasted effort if the tool isn't accurate:

\begin{tcolorbox}[thematicquote]
\itshape``So I think that worked really well, especially with larger [PRs] [...] you can at least use it if you trust it.'' \hfill [P7]
\end{tcolorbox}

\begin{tcolorbox}[thematicquote]
\itshape``I could go on for hours, just to realize I can never do this. Then I’ve just lost a few hours trying to pursue something that wasn’t possible.'' \hfill [P5]
\end{tcolorbox}

\subsubsection{Efficiency and Thoroughness}

A strong theme was the AI’s impact on code review efficiency and the thoroughness or quality of reviews. Developers reported that the AI assistant could \textbf{speed up the review process, reduce reviewers’ workload on tedious tasks, and potentially catch more issues}, although some participants remarked that it \textbf{sometimes focuses on low-priority findings, introducing noise}.

Beyond efficiency, participants also pointed to improvements in \textbf{review quality}. Some noted that the assistant could identify issues that might otherwise go unnoticed:

\begin{tcolorbox}[thematicquote]
\itshape``There are probably findings you get in the report that you don't find when you do it manually.'' \hfill [P10]
\end{tcolorbox}

Additionally, interviewers felt that the assistant would be particularly helpful on \textbf{large pull requests} where it would be difficult for a human reviewer to catch everything:

\begin{tcolorbox}[thematicquote]
\itshape``It’s taken someone two weeks to write it [...] giving it 15 minutes, you won’t have a chance to understand it, at least not well enough to find the hard stuff. I would imagine that the tool would actually raise a flag for a potential deadlock or race condition as well.'' \hfill [P7]
\end{tcolorbox}

For some participants, the assistant reduced the effort involved in reviewing by \textbf{removing the need to search through the codebase or external documentation}:

\begin{tcolorbox}[thematicquote]
\itshape``But I really like the integration with the requirements part, because if I open up [a PR] and I don't know what it's about. [...] The first thing I do every time is I open the [Jira] ticket anyway, because I need to see what is supposed to have been achieved here. So I think that's a really nice functionality to have'' \hfill [P3]
\end{tcolorbox}

However, some participants noted that the assistant occasionally surfaced low-priority or unclear findings. This was also observed during review sessions, where participants sometimes expressed \textbf{difficulty distinguishing important findings from minor ones, especially in lengthy summaries generated by the assistant}.

\subsubsection{Design Expectations and Limitations}

Many participants shared expectations about how an AI assistant for code review should behave and be integrated. A recurring theme was the desire for \textbf{seamless integration into existing workflows and tools}. Rather than switching to a new interface, several participants expressed that it would be preferable to access the assistant directly from familiar environments like GitHub, Slack, or their IDEs. As one participant put it:

\begin{tcolorbox}[thematicquote]
\itshape``I think most of the developers don't want to use something new, like a [new] UI, but rather have an integration to what exists.'' \hfill [P11]
\end{tcolorbox}

Building on this, another participant described a preference for having the assistant's comments embedded directly into GitHub's interface, with expandable in-line comment boxes, while also having the ability to further ask questions in a chat interface. Yet another participant described how having it in the pipeline through a Slack bot could be useful.

\begin{tcolorbox}[thematicquote]
\itshape``Maybe a Slack auto bot could even be triggered on each message [...] and a full review could be dropped as a message under that thread.'' \hfill [P11]
\end{tcolorbox}

Beyond integration, participants also critiqued how the LLM assistant's output was presented. Several interviewees felt that the \textbf{LLM feedback was overly long or difficult to scan}. One participant noted:

\begin{tcolorbox}[thematicquote]
\itshape``I mean, what's important is that it very clearly lists the file. I'd rather have it list the file and the line number and be very specific, in a short way.'' \hfill [P5]
\end{tcolorbox}

\textbf{Response time} was another point of friction. While some delays were tolerated, long wait times were cited as a major barrier to adopting the tool in real development workflows:

\begin{tcolorbox}[thematicquote]
\itshape``I think the speed and accuracy are mainly what need to be improved. [...] I wouldn't use this if it took, I don't know, how many minutes it took for it to respond.'' \hfill [P3]
\end{tcolorbox}

One participant reflected on whether the quality of interaction also depended on their \textbf{own ability to ask good questions}:

\begin{tcolorbox}[thematicquote]
\itshape``Maybe my way of asking questions was also wrong. I did not always feel like I got the response that I was asking for. So, I might need to learn how to be more detailed in my questions.'' \hfill [P10]
\end{tcolorbox}

Many limitations were traced to a \textbf{lack of access to broader context, such as architectural documentation, internal conventions, or metadata}:

\begin{tcolorbox}[thematicquote]
\itshape``Ideally, you want to inject as much relevant information as possible [...] like the JIRA ticket, relevant [documentation] pages, the codebase itself, the README, and any similarly named repositories that might be connected to the same service.'' \hfill [P5]
\end{tcolorbox}

Some participants also highlighted that the \textbf{LLM-assistant's usefulness depended in part on how good the documentation and PR descriptions are to begin with}. During the testing, one reviewer reflected that the tool could help keep the documentation up-to-date by suggesting updates that align with PR changes.

\vspace{1mm}
\subsubsection{Usage Contexts and Interaction Patterns}

Participants expressed a range of preferences, strategies, and situational factors that shaped how they interacted with the AI assistant, often shaped by the review context. These patterns included both the predefined interaction modes: Mode A (Co-Reviewer) and Mode B (Interactive Assistant), as well as emergent workflows that blended or extended beyond them.

Many participants \textbf{found Mode A (Co-Reviewer) especially helpful for getting oriented in a pull request}. They described the high-level summaries and suggestions provided at the beginning of the review as useful for gaining quick context, particularly in unfamiliar or complex codebases:

\begin{tcolorbox}[thematicquote]
\itshape``I prefer this one [Mode A] where you actually get the overview directly [...] it had a lot of good pointers, that it already found.'' \hfill [P12]
\end{tcolorbox}

\begin{tcolorbox}[thematicquote]
\itshape``The first engine [Mode A] that gave me a breakdown of everything [...] that was quite clever, and I would gladly use that.'' \hfill [P7]
\end{tcolorbox}

Mode A was also described as particularly \textbf{useful for low-risk PRs}:

\begin{tcolorbox}[thematicquote]
\itshape``Let's say the change is relatively small and it's not causing any risk, then I would definitely go with the first one [Mode A], where I let AI do most of the work.'' \hfill [P11]
\end{tcolorbox}

Many participants saw the assistant, especially mode A, as \textbf{valuable for newcomers}:

\begin{tcolorbox}[thematicquote]
\itshape``I think if I were in a new team, and I am unsure what is happening, then it could be really good to start with a summary.'' \hfill [P8]
\end{tcolorbox}

\begin{tcolorbox}[thematicquote]
\itshape``Yeah, if you can write questions like 'What is this?' or 'What is this really about?', it could also be a very good tool to get to know the codebase and to learn as a new guy.'' \hfill [P9]
\end{tcolorbox}

Additionally, some saw mode A as \textbf{useful in teams where code review standards are lower or in teams that prefer other methods for reviewing code, such as pair programming}. In such cases, the assistant would serve as a fallback mechanism.

\begin{tcolorbox}[thematicquote]
\itshape``I think it would be great for those who usually just skim through and say, 'It looks good to me'. [...] I think the biggest effect would be for those developers, I guess, and those teams.'' \hfill [P5]
\end{tcolorbox}

Mode B was less preferred in general, but some participants remarked that in cases where they are \textbf{already familiar with the codebase or if they wanted to maintain full control over the review process, they would prefer Mode B}:

\begin{tcolorbox}[thematicquote]
\itshape``But if it's in some codebase I already know, some codebase where we have a lot of experience and have worked in it a lot. It could probably be nice to have [Mode B].'' \hfill [P8]
\end{tcolorbox}

In some cases, participants preferred a \textbf{combination of both modes} or expressed that the preferred interaction mode depended on the situation:

\begin{tcolorbox}[thematicquote]
\itshape``I think I'm 50/50 [...] both are useful. One is on demand, the other one is on its own.'' \hfill [P1]
\end{tcolorbox}

Several participants \textbf{proposed additional usage patterns} that were not strictly defined by the study design. For instance, some saw mode A as useful for the author before submitting the PR, rather than during review:

\begin{tcolorbox}[thematicquote]
\itshape``I feel it might not be as much of a review help. I think it might be a pre-review help.'' \hfill [P10]
\end{tcolorbox}

Others \textbf{proposed an alternative interaction mode} where engineers conducted a human-led review first and then used the assistant to validate or catch anything they might have missed:

\begin{tcolorbox}[thematicquote]
\itshape``You start out with it just to sum up what the code is doing. Then I look for issues, and then I can ask, 'Are there any further issues?'' \hfill [P3]
\end{tcolorbox}

Participants also frequently noted that Mode A was especially \textbf{helpful for large PRs}:

\begin{tcolorbox}[thematicquote]
\itshape``Especially for large PRs, it's nice to get the breakdown on what's happening [...] because usually, you always have to do that sort of manually anyway.'' \hfill [P3]
\end{tcolorbox}

However, there was also some uncertainty about how \textbf{effective the assistant would be at scale}. One participant expressed concerns about the assistant’s ability to handle very large codebases or complex business logic:

\begin{tcolorbox}[thematicquote]
\itshape``I think it's going to be a bottleneck for such things, because there will be so many moving parts in it, so much business logic going around.'' \hfill [P1]
\end{tcolorbox}

\section{Implications}

Our study yields several practical implications for integrating LLMs into code review workflows. First, AI assistance should be embedded within developers’ existing tools (such as GitHub, GitLab, IDEs, or Slack) to minimize friction and support natural adoption. Second, output must be concise, well-structured, and actionable, prioritizing critical findings with precise references to affected files and lines. Fast response times are essential to preserve reviewer flow, although more comprehensive agentic reviews may be acceptable when integrated into automated pipelines. Supporting both proactive (AI-led summaries) and reactive (on-demand Q\&A) modes of interaction is key, with a general developer preference for AI-led summaries in large or unfamiliar pull requests. For meaningful support, the assistant must be context-aware and have access to relevant information such as code diffs, source files, and requirement documents. While our tool addressed this via a retrieval-augmented setup, participants still highlighted the need for deeper contextual integration. Finally, an LLM-enabled assistant also shows promise as a pre-review aid, helping authors catch simple issues before submitting a pull request, thus improving code quality upstream in the development lifecycle.

\section{Concluding Remarks}

This paper presented a field study and field experiment conducted at WirelessCar to explore the integration of Large Language Models (LLMs) into real-world code review workflows. The study surfaces persistent challenges in current review practices, such as context switching, reviewer fatigue, inconsistent review depth, and developer perceptions of how LLMs can augment the process. By evaluating two interaction modes (AI-led reviews and on-demand assistance), we found that developers generally value AI-generated summaries and contextual clarifications, particularly in large or unfamiliar pull requests. However, concerns around trust, false positives, response latency, and integration friction remain. While most participants preferred the AI-led mode in unfamiliar or low-risk scenarios, preferences were context-dependent, with some favoring human-led reviews when code familiarity or criticality increased.

This study contributes practical insights into how LLMs can complement human reviewers, rather than replace them. Our findings suggest a promising path forward: integrating AI assistance more tightly into existing development environments, improving response quality and speed, and offering adaptive interaction modes tailored to developer needs.

\bibliographystyle{IEEEtran}
%\bibliography{ref}

% Generated by IEEEtran.bst, version: 1.14 (2015/08/26)

\end{document}